\documentclass[acmtog]{acmart}
\acmSubmissionID{118}

\setcitestyle{nosort,square} %

\usepackage[ruled]{algorithm2e} %

\usepackage{pifont}
\usepackage{colortbl}
\usepackage{xcolor}
\usepackage{float}

\SetAlFnt{\small}
\SetAlCapFnt{\small}
\SetAlCapNameFnt{\small}
\SetAlCapHSkip{0pt}

\newcommand{\cmark}{\ding{51}}%
\newcommand{\xmark}{\ding{55}}%

\newcommand{\greencheck}{{\color{green}\cmark}\xspace}
\newcommand{\redcheck}{{\color{red}\xmark}\xspace}
\newcommand{\yellowcheck}{{\color{yellow}(\cmark)}\xspace}
\newcommand{\yellow}{\cellcolor{yellow!10}\yellowcheck}
\newcommand{\green}{\cellcolor{green!12.5}\greencheck}
\newcommand{\red}{\cellcolor{red!12.5}\redcheck}

\setcopyright{none}
\renewcommand\footnotetextcopyrightpermission[1]{}
\acmJournal{TOG}

\begin{document}
\title{Neural Lithography: Close the Design-to-Manufacturing Gap in Computational Optics with a 'Real2Sim' Learned Photolithography Simulator}

\begin{teaserfigure}
\centering
\includegraphics[width=0.95\linewidth]{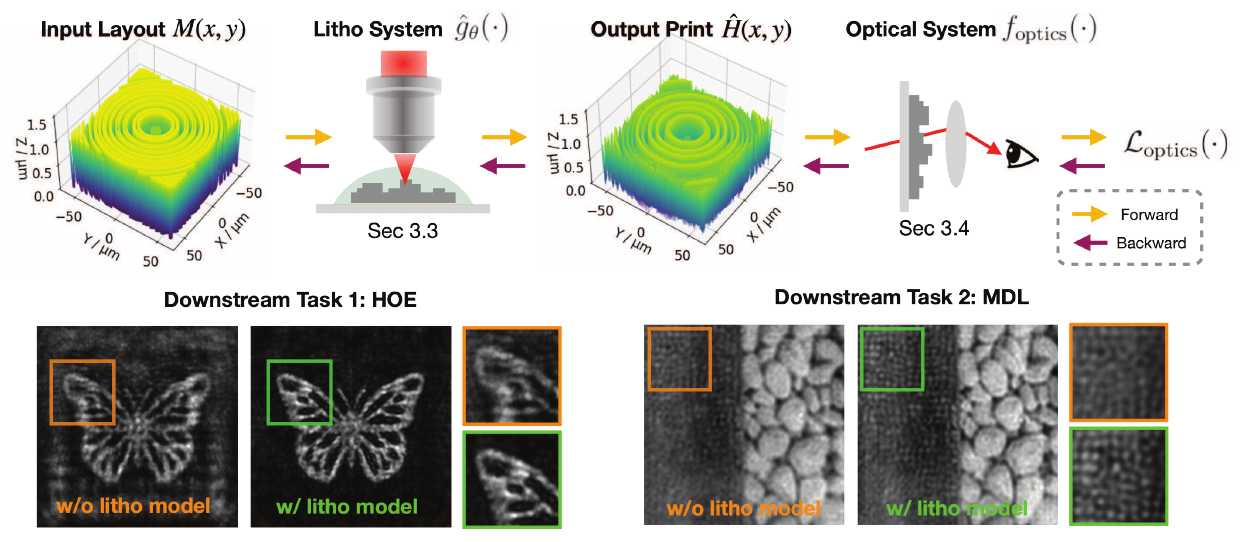}
\caption{\label{schema}\textbf{Our neural lithography framework for improving the image quality of fabricated optics.} We digitalize a real-world photolithography system via learning its digital twin $\hat{g}_{\theta}$ (Sec.~\ref{subsec: litho simulator}) from real measurements. This helps in revealing fabrication constraints during the downstream optical design tasks, including the task-specific light transport $f_\text{optics}$ and metric $\mathcal{L}_\text{optics}$ (Sec.~\ref{subsec: downstream tasks}).
Our work showcases an improved performance on the design of holographic optical elements (HOE) and multilevel diffractive lenses (MDL) with our differentiable neural lithography simulator in the design loop.
}
\end{teaserfigure}

\author{Cheng Zheng} %
\authornote{Corresponding author.}
\authornote{Equal contribution.}
\orcid{0009-0001-0367-8701}
\affiliation{%
 \institution{Massachusetts Institute of Technology}
 \streetaddress{77 Massachusetts Ave}
 \city{Cambridge}
 \state{MA}
 \postcode{02139}
 \country{USA}}
\email{chengzh@mit.edu}
\author{Guangyuan Zhao}
\authornotemark[1]
\authornotemark[2]

\orcid{0009-0008-7822-9565}
\affiliation{%
 \institution{The Chinese University of Hong Kong}
 \city{Hong Kong}
 \country{China}
}
\email{zhaoguangyuan2021@gmail.com}
\author{Peter T.C. So}
\orcid{0000-0003-4698-6488}
\affiliation{%
\institution{Massachusetts Institute of Technology}
\streetaddress{77 Massachusetts Ave}
\city{Cambridge}
\state{MA}
\country{USA}}
\email{ptso@mit.edu}
\renewcommand\shortauthors{Zheng, \textit{et al.}}

\begin{abstract}
   We introduce neural lithography to address the ‘design-to-manufacturing’ gap in computational optics. Computational optics with large design degrees of freedom enable advanced functionalities and performance beyond traditional optics. However, the existing design approaches often overlook the numerical modeling of the manufacturing process, which can result in significant performance deviation between the design and the fabricated optics. To bridge this gap, we, for the first time, propose a fully differentiable design framework that integrates a pre-trained photolithography simulator into the model-based optical design loop. Leveraging a blend of physics-informed modeling and data-driven training using experimentally collected datasets, our photolithography simulator serves as a regularizer on fabrication feasibility during design, compensating for structure discrepancies introduced in the lithography process. We demonstrate the effectiveness of our approach through two typical tasks in computational optics, where we design and fabricate a holographic optical element (HOE) and a multi-level diffractive lens (MDL) using a two-photon lithography system, showcasing improved optical performance on the task-specific metrics. The source code for this work is available on the project page: \textcolor{magenta}{\url{https://neural-litho.github.io}}.

\end{abstract}




\maketitle
\fancyfoot{}
\thispagestyle{empty}

\section{Introduction}

In recent years, we have witnessed a revolution in the design of computational optical elements, extending well beyond conventional lenses and mirrors. 
Holographic optical elements, as a representative example, have found applications in various domains, including 3D imaging as speckle generators~\cite{Inteld415}, augmented reality as waveguide combiners~\cite{jang2020design, jang2022waveguide}, laser scanning as beam splitters~\cite{hahn2020rapid}, and others~\cite{kuo2020high, blachowicz2021optical}. 
Recent advances in deep learning and auto-differentiation~\cite{baydin2018automatic} have further enabled the development of end-to-end design pipelines for diffractive-~\cite{sitzmann2018end} and meta-optical~\cite{tseng2021neural} imaging systems, demonstrating promising results in numerous imaging tasks, such as depth from defocus~\cite{ikoma2021depth}, hyperspectral imaging~\cite{baek2021single}, high-dynamic-range imaging~\cite{metzler2020deep, sun2020learning}, and image pixel super-resolution~\cite{sitzmann2018end, sun2020end}. 
These computational imaging devices facilitate the creation of compact, miniaturized, and multifunctional imaging devices and expand the capabilities of conventional vision systems, offering unprecedented opportunities to meet diverse imaging requirements.

Nonetheless, a crucial challenge remains: fabricating elements that fulfill design objectives, offering large design degrees of freedom (DOF), and being cost-effective. Present popular fabrication techniques include e-beam lithography~\cite{tseng2021neural}, diamond tuning~\cite{sitzmann2018end}, and photolithography~\cite{levin2013fabricating, tan2021codedstereo, 
boominathan2020phlatcam, orange20213d, baek2021single} . Among these techniques, photolithography, which is the focus of this work, uniquely allows 3D manufacturing capacity~\cite{blachowicz2021optical, you2020mitigating}, exhibits low cost, easy prototyping, and allows larger and flexible design space~\cite{lee2023design}, which are more or less unmet in other fabrication methods.
Yet, a critical step, the numerical modeling of the photolithography process in the design loop, is often overlooked in previous studies ~\cite{jang2020design, tan2021codedstereo, boominathan2020phlatcam}.

We identify this deficiency, a phenomenon we term the \textbf{"design-to-manufacturing gap" in computational optics}, as a fundamental issue in the current design and fabrication process. This gap, arising from differences between the theoretical and actual outputs of the lithography system due to light diffraction and photo-chemical processes, has significant implications~\cite{wang2020toward,roques2022toward}. This impairs not only the accuracy of optical information processing, the diffraction efficiency~\cite{ikoma2021depth, baek2021single} but also the minimum controllable feature size, which is a critical parameter in computational optics as it determines the maximum diffraction angle~\cite{levin2013fabricating} and the étendue of the optical system~\cite{kuo2020high}. While an intuitive solution is to numerically incorporate the manufacturability during the task design process, developing a high-fidelity simulator to ensure flawless integration and efficiently co-optimizing the task-specific design process remains a further challenge.

In response to these challenges, we propose a framework that integrates a pre-trained neural lithography simulator into the model-based optical design process to collectively work with the task-related computational optics models to let the design meet task-specific metrics while maintaining fabrication feasibility. This simulator, trained from the dataset explored in a real-world lithography system using a synergistic blend of traditional physics-based and data-driven modeling, compensates for the deviation between the ideal and the real lithography process. In this way, our framework mitigates the "design-to-manufacturing gap" and enhances the fabrication fidelity of the design, leading to superior performance in downstream computational optical tasks. 

Specifically, our contributions are:
\begin{itemize}
    \item A "real2sim" methodology that learns high-fidelity neural simulators that accurately model real-world photolithography systems from 2.5D atomic force microscope (AFM) measurements.
    \item A design and fabrication co-optimization pipeline that models the downstream computational optics task and photolithography process as two intersected differentiable simulators to optimize for desired optical performance post-manufacturing, using inline holography and imaging as example tasks.
    \item Experimental verification of improved optical performance by fabricating optimized diffractive optical elements with a two-photon lithography (TPL) system and evaluating on a home-built optical setup.
\end{itemize}

{\paragraph{Scope} This research employs a 'Real2Sim' learning approach to construct high-fidelity neural photolithography simulators, specifically validated on TPL systems using AFM datasets. We take initial steps in the broad domain of co-optimizing manufacturability with computational optics design. We believe that the insights provided by our work are invaluable and could inspire further research in the fields of computational optics and photolithography.
}

\section{Related work}

\textbf{Model-based optical proximity correction (OPC) and inverse lithography technology (ILT)} are computational lithography techniques commonly used to compensate for errors due to diffraction or process effects during or after fabrication. Both methods are based on a simulator of the lithography system. The former refers to the solution of perturbing surrounding pixels of input design geometry through feedback from the simulator. The latter formulates the lithography enhancement task as an inverse design problem and sets the input design geometry as the targets with no limitation on the design DOF~\cite{cecil2022advances}. Early research built a physics-based simulator of the lithography system and solved the ILT problem with adjoint methods/gradient descent/backpropagation~\cite{cecil2022advances,poonawala2007mask}. Recently, researchers have been using deep learning methods, such as LithoGAN~\cite{ye2019lithogan}, GAN-OPC~\cite{yang2018gan}, and optics-inspired neural networks~\cite{yang2022generic, yang2022large}, neural-ILT~\cite{jiang2020neural} to speed up the OPC or ILT process through learning a fast proxy from expensive simulated data generated from the physics-based simulator. 

Our work addresses a critical challenge in computational lithography: constructing a high-fidelity digital twin of real-world lithography systems. Such systems have been modeled in a white-box manner, building physics-based simulators approximated from first principles in optics and photochemistry, like the Hopkins diffraction model for Kohler illumination and sigmoid functions for resist photochemical reactions~\cite{cobb1996mathematical, poonawala2007mask, chevalier2021rigorous}. However, they fall short in adapting to system-specific and photoresist-specific variations~\cite{arthur1996investigation}. These limitations are not fully remedied even for industry standards like Sentaurus Lithography, which has evolved to simulate 3D resist profiles~\cite{Synopsys2023SLitho}.

Against this backdrop, our research, while rooted in a learned representation of the lithography system, introduces a paradigm-shifting contribution that may significantly improve lithography performance. We employ a 'real2sim' methodology~\footnote{real2sim is an emerging concept in robotics which advocates for building high-fidelity simulators directly from real-world data~\cite{real2simICRA2023}.} for training our differentiable simulator from real 2.5D layout-print pairs, in contrast with the conventional 'sim2sim' method. Therefore, our method offers a more adaptable and accurate way to digitalize the photolithography system while de-emphasizing the computational speed. Furthermore, our focus on 3D structures necessitates a more informative dataset – a '2.5D' heightmap of prints instead of '2D' datasets for common edge position error (EPE) corrections~\cite{cecil2022advances}.

\noindent\textbf{Optimization of photolithography for micro/nano optics} has seen recent advancements (Table~\ref{tab: comparison of optimization work}). Wang et al. utilized grid search for Dammann gratings fabrication~\cite{wang2020toward}, while Lang et al. implemented a parameterized physics model for multi-photon lithography~\cite{lang2022towards}. Chevalier et al. integrated a physics-based lithography model to optimize microlens array fabrication using atomic force microscopy data~\cite{chevalier2021rigorous, giessibl2003advances}.
Of particular relevance is Liao et al.'s work~\cite{liao2022line}, which leveraged machine learning to construct an i-line lithography simulator from real 2D binary scanning electron microscopy (SEM) data, leading to a non-differentiable OPC to optimize for the EPE. However, this process presents computational challenges, and the limited SEM dataset information hinders complex design capabilities.

\begin{table}[ht]
\fontsize{7pt}{7pt}\selectfont
    \centering
    \begin{tabular}{c*{5}{|m{1cm}}}
     \toprule
       & \textbf{~\cite{wang2020toward}}  & \textbf {~\cite{chevalier2021rigorous}} &\textbf{~\cite{lang2022towards}} & \textbf {~\cite{liao2022line}} & \textbf{Ours} \\\midrule
       \rule{0pt}{8pt}
       Data-driven &\hfil\yellow &\hfil\red &\hfil\green &\hfil\green &\hfil\green \\
      \rule{0pt}{8pt}
       Physics & \hfil\yellow &\hfil\green &\hfil\green &\hfil\yellow &\hfil\green\\
       \rule{0pt}{8pt}
       2.5D  & \hfil\green & \hfil\green &\hfil\green &\hfil\red &\hfil\green\\
       \rule{0pt}{8pt}
       Nanoscale & \hfil\green &\hfil\green &\hfil\red &\hfil\green &\hfil\green \\
       \rule{0pt}{8pt}
       End-to-end$\nabla$ & \hfil\red &\hfil\red &\hfil\red &\hfil\red &\hfil\green \\
       \bottomrule
    \end{tabular}
    \caption{\textbf{Comparison of photolithography optimization approaches on fabricating the micro-/nano-optics} along the axes of data-driven modeling, physics-aware modeling, 2.5D optimization capability, nanoscale precision, and end-to-end differentiability.}
    \label{tab: comparison of optimization work}
\end{table}
\vspace{-2em}

In contrast, our work employs a dataset measured by AFM, enabling optimization for $2.5D$ geometry of computational optics. Furthermore, our approach is fully differentiable, combining optical design and fabrication correction end-to-end. By incorporating fabrication constraints directly in the design loop, this methodology navigates complex design space more effectively.

\noindent\textbf{Differentiable optics} refers to an emerging technique in computational optics that models an optical system with chained differentiable functions to design optical systems using gradient-based optimization methods, which is computationally efficient and effective in exploring large and complex design space. Recently, based on differentiable optics, a large body of work in computational imaging has studied the joint design of the optics and the reconstruction algorithms~\cite{sitzmann2018end, tseng2021neural} using the physics-based (white-box) simulators. 

\begin{figure}[h!]
\centering
\includegraphics[width=0.98\linewidth]{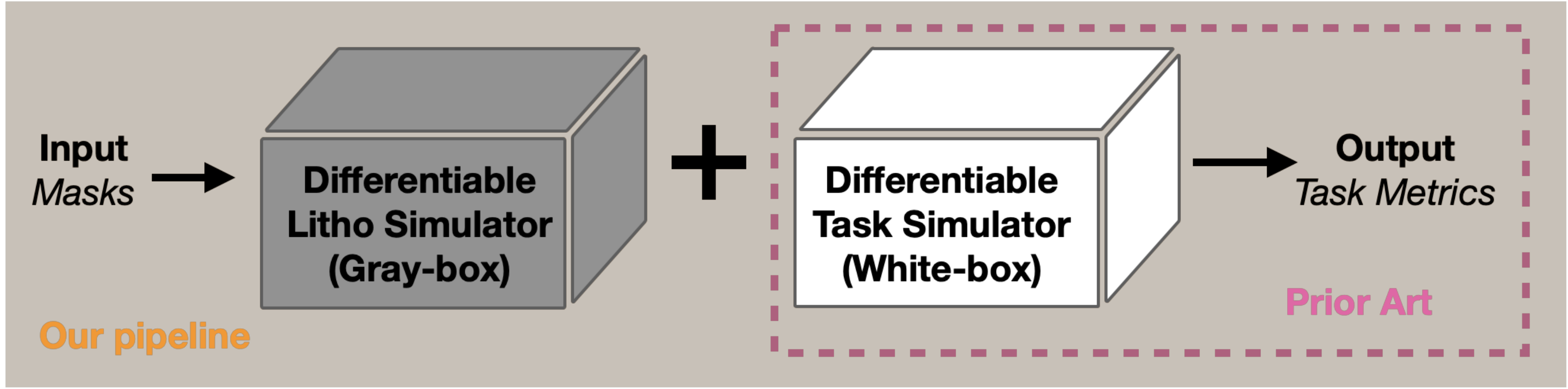}
\caption{\textbf{Our work advances the differentiable optics via chaining the prior art of white-box design with a gray-box lithography simulator.} 
The task simulator uses the first principles of optics to build the task light transport. We add a 'real2sim' learned gray-box lithography simulator in the design loop. 
}
\label{fig: chained_simulators}
\end{figure}

Our neural/differentiable lithography also lies in the regime of differentiable optics. The uniqueness is that our work is the first to integrate design and manufacturing as two differentiable modules of an end-to-end optimization pipeline for high-performance computational optics tasks (Fig.~\ref{fig: chained_simulators}). We, on the one hand, utilize the differentiable optics-based first principle to enable the optical design and, on the other hand, build a differentiable neural lithography simulator in a 'real2sim' manner (the gray-box simulator) for the lithography process. The latter resembles very recent work on 'hardware-in-the-loop' optimization of camera ISP~\cite{tseng2019hyperparameter} and holographic display system~\cite{peng2020neural, chakravarthula2020learned}.

\section{Methodology}
We formulate the problem of the fabrication-embedded design of computational optics in subsection~\ref{subsec: prob stat}. We then detail our solution of using model-based optimization for jointly optimizing the fabrication and the design process in subsection~\ref{subsec: mbo}. We illustrate our protocol for digitalizing the lithography system in subsection~\ref{subsec: litho simulator}. Next, we present two tasks of designing and fabricating holographic optical element (HOE) and multi-level diffractive lens (MDL) in subsection~\ref{subsec: downstream tasks}. 

\subsection{Problem statement} \label{subsec: prob stat}
We are interested in finding a mask layout $M(x,y)$ to feed into a photolithography system, such that the output fabricated structure being used as an optical element with height profile $H(x,y) = g(M(x,y))$ achieves the best performance in the downstream computational optics tasks, where $g$ denotes the underlying mathematical representation of the fabrication process. The whole pipeline is described as follows:

\begin{equation}
        M^\star\left(x,y\right) = \operatorname*{arg\,min}_{M\left(x,y\right)} \mathcal{L}_{\text{optics}}(f_{\text{optics}}( g(M\left(x,y\right)))).
        \label{eq: obj computational optics}
\end{equation} 
Without loss of generality, we use $f_{\text{optics}}$ and $\mathcal{L}_{\text{optics}}$ to represent the task-specific forward light transport and loss function of the downstream tasks, respectively. 

The manufacturing process, denoted by $g$, is pivotal for the overall pipeline's efficacy. While past research in computational optics often views this process as an identical mapping, $g(M(x,y))=M(x,y)$~\cite{tan2021codedstereo, boominathan2020phlatcam, jiang2020neural}, this assumption falls short for features around the diffraction limit. In the photolithography process, the light's diffraction, system aberrations, and chemical reactions cause deviations from the input layout, negatively impacting the performance of the fabricated computational optics, which depends significantly on precise structure geometry. We focus on creating computational optics with desired optical performance post-manufacturing.

\begin{figure}[h!]
\centering
\includegraphics[width=0.9\linewidth]{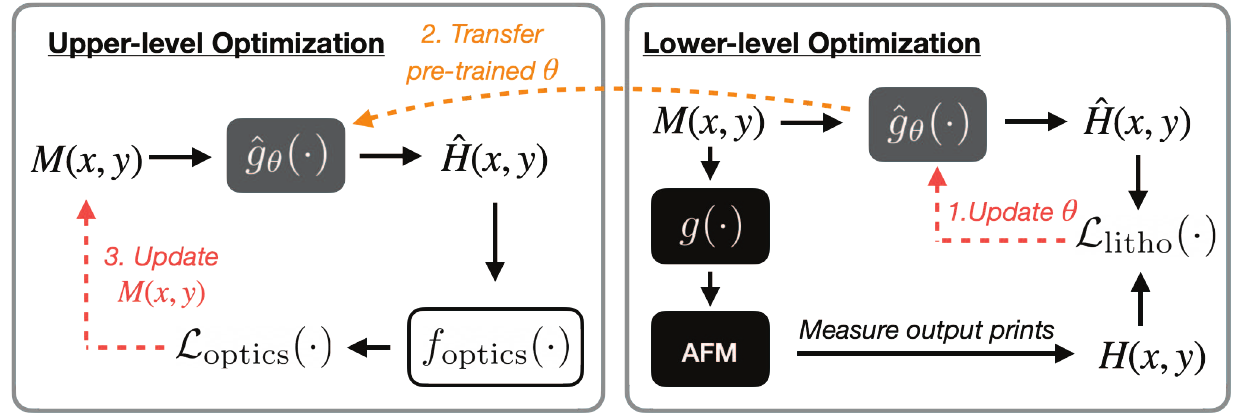}
\caption{\textbf{Bilevel optimization in neural lithography.} 
The upper-level optimization optimizes the mask $M(x,y)$ while the lower-level optimization optimizes the litho digital twin $\hat{g}_{\theta}$, the result of which is imported and fixed as a constraint in the upper-level optimization. 
}
\label{fig: bilevel optim}
\end{figure}

\subsection{Model-based differentiable optimization of computational optics with a lithography simulator} \label{subsec: mbo}
Our solution reformulates the objective in Eq.~\ref{eq: obj computational optics} as a bi-level optimization problem in Fig.~\ref{fig: bilevel optim} ~\cite{colson2007overview}, structured as:

\textbf{Upper-level (Leader) optimization (subsec.~\ref{subsec: downstream tasks}):} 
\textit{Optimize the performance of the computational optics in the specific task based on predictions from the learned lithography simulator (e.g., minimize wavefront error, maximize optical efficiency, or achieve desired functionality).} The objective function in Eq.~\ref{eq: obj computational optics} is rewritten as:

\begin{equation}
        M^\star\left(x,y\right) = \operatorname*{arg\,min}_{M\left(x,y\right)} \mathcal{L_{\text{optics}}}(f_{\text{optics}}(\hat{g}_{\theta}(M\left(x,y\right)))),
        \label{eq: rewrite obj computational optics}
\end{equation}
where we substitute the unknown underlying lithography system $g$ in Eq.~\ref{eq: obj computational optics} with $\hat{g}_{\theta}$, a digital twin of $g$ with parameters $\theta$ learned in the lower-level optimization.

\textbf{Lower-level (Follower) optimization (subsec.~\ref{subsec: litho simulator}):} \textit{Pre-train a lithography digital twin $\hat{g}_{\theta}$ that accurately predicts height profile $H(x,y)$ of the printed structure given input layout $M(x,y)$}. The optimization process is minimizing the difference between the lithography simulator prediction $\hat{H}(x,y) = \hat{g}_{\theta}(M(x,y))$, and the experimentally measured $2.5\text{D}$ continuous height $H(x,y)$:

\begin{equation} \label{eq: obj real}
    {\theta}^{\star} = \operatorname*{arg\,min}_{{\theta}} \mathcal{L}_{\text{litho}}(H(x,y), \hat{H}(x,y)),
\end{equation} 
where $\mathcal{L}_{\text{litho}}$ denotes the loss function for this supervised learning, and here we use the mean absolute error.

\textbf{To summarize:} The leader problem optimizes computational optical element design using the prediction from the neural lithography simulator, learned independently in the follower problem. These interconnected problems, addressed in a bilevel optimization framework, close the design-to-manufacturing gap with robust designs and better optical performance.

\subsection{'Real2Sim' learning a neural lithography simulator}\label{subsec: litho simulator} 

\subsubsection{Lithography forward model.}\label{subsec: litho fwd model} 
We construct a physics-based neural network $\hat{g}_{\theta}$ as the differentiable digital twin of the lithography system $g$, referred to as the physics-based learning (PBL) model in this paper. The modeling of the photolithography process normally consists of an optical model and a resist model~\cite{cobb1996mathematical, chevalier2021rigorous, lang2022towards}:
\begin{itemize}
    \item The \textbf{optical model} calculates the light exposure dosage distribution on the photoresist, leading to the aerial image; 
    \item The \textbf{resist model} models the photo-chemical reaction and development processes, leading to the final resist profile.
\end{itemize}
 
We first illustrate $\hat{g}_{\theta}$ in the context of the general photolithography model and then detail the specific arrangements of the TPL system used in this work. In our context, \textbf{the resist profile is the ultimate fabricated structure}, as it does not undergo further processing steps. Inspired by the work from~\cite{tseng2022neural} on photo-finishing, we also apply the pointwise and areawise neural networks inside $\hat{g}_{\theta}$ to model the corresponding operations, respectively (see details in Supplements). Anatomy of the model $\hat{g}_{\theta}$ is visualized in Fig.~\ref{fig: litho digital twin} and details of the sub-models are:

\begin{figure}[h!]
\centering
\includegraphics[width=0.95\linewidth]{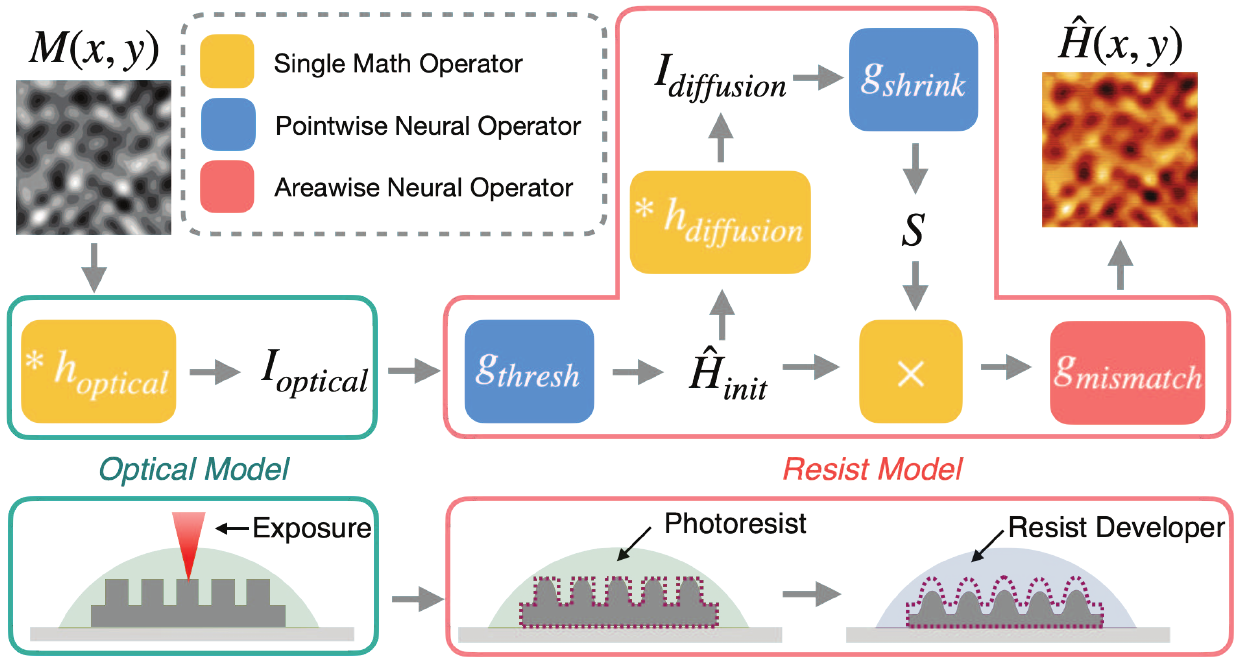}
\caption{\textbf{Anatomy of our litho digital twin $\hat{g}_{\theta}: M(x,y)\rightarrow \hat{H}(x,y)$.} The input layout $M(x,y)$ goes through the optical model (exposure process) and resist model(photochemical reaction and development processes). The litho digital twin predicts the final fabricated structure height as $\hat{H}(x,y)$.}
\label{fig: litho digital twin}
\end{figure}

The \textbf{optical model} $g_{\text{illum}}$ transforms the input layout mask $M(x,y)$ into an exposure dosage distribution, also known as aerial image, $I_{\text{aerial}}(x,y) = g_{\text{illum}}(M(x,y))$. Depending on the illumination sources, $g_{\text{illum}}$ normally incorporate illumination kernel(s), such as a Gaussian point spread function (PSF) under point-source illumination ~\cite{auzinger2018computational,saha2017effect} or a series of convolution kernels decomposed from transmission-cross coefficient (TCC) matrix in Hopkin's formulation under Köhler’s illumination~\cite{cobb1996mathematical, liao2022line}. Specifically, in the point-scanning TPL system used in this work~\cite{harinarayana2021two}, $g_{\text{illum}}$ is modeled as convolution between the mask $M$ and a Gaussian PSF $h_{\text{optical}}$ with a hyper-parameter $\sigma_{\text{optical}}$ being its half-width at half maximum.

The \textbf{resist model} computes the photochemical reaction during exposure and accounts for the structure deformation during development. \textit{The initialization of photochemical reaction process} is modeled with a pointwise neural network $g_\text{thresh}$ to accurately handle the pixel-wise thresholding operation, resulting in the initial resist profile $\hat{H}_{\text{init}}(x,y) = g_{\text{thresh}}(I_{\text{aerial}}(x,y))$. Since the \textit{diffusion of reactants}  introduces cross-talk between nearby printed features, we use a Gaussian kernel $h_{\text{diffusion}}$ with a learnable standard deviation $\sigma_{\text{diffusion}}$ to approximate the spatial extent of diffusion~\cite{lang2022towards}. As a result, at the end of the reaction we have:
\begin{equation}
I_{\text{diffusion}}(x,y) = \left(\hat{H}_{\text{init}}*h_{\text{diffusion}}\right)\left(x,y\right).
\end{equation}

\textit{The development process} introduces extra deformation to the printed structure, which has been observed to be unidirectional shrinkage perpendicular to the substrate due to the dissolution of soluble components and the capillary forces during the drying process\cite{meisel2006shrinkage, d2021modeling}. We model this anisotropic shrinkage $S$ with an operation $g_\text{shrink}$ represented by another pointwise neural network, giving $S(x,y) = g_\text{shrink}(I_\text{diffusion}(x,y))$. Thus, the resist profile is the product of $\hat{H}_\text{init}$ and $S$ after development. 

To further mitigate mismatch from the above approximations, we introduce an areawise neural network as a learnable operator $g_\text{mismatch}$, giving the final fabricated structure as:
\begin{equation}
    \hat{H}(x,y) = g_{\text{mismatch}}((\hat{H}_{\text{init}}\cdot S)(x,y)).
\end{equation}

\textbf{Chaining the sub-models}, we get the predicted height profile of the fabricated structure as: 

\begin{subequations}
   \begin{align}
    \hat{H}(x,y) &= \hat{g}_{\theta}(M(x,y)),  \\
    &= g_\text{mismtach}\left(\hat{H}_{\text{init}}\cdot g_{\text{shrink}}(\hat{H}_{\text{init}}*h_{\text{diffusion}})\right),\\
    \hat{H}_{\text{init}}(x,y) &= g_{\text{thresh}}\left(g_{\text{illum}}(M(x,y))\right).
    \end{align} \label{eq: diff modules of system}
\end{subequations}

When applying the 'real2sim' pipeline to diverse photolithography processes, one cannot rely on a 'one-size-fits-all' model. To effectively accommodate system-specific or process-dependent variations, hyper-parameters or operations must be fine-tuned. Key among these are $h_{\text{optical}}$ or $g_{\text{illum}}$ (pertaining to the illumination system) and $h_{\text{diffusion}}$ (associated with the photoresist and its processing protocol). More details about the model architecture and the training process are in the Supplements.

\subsubsection{Constructing dataset for learning a lithography simulator} \label{subsec: dataset}

\begin{figure}[ht]
\centering
\includegraphics[width=0.9\linewidth]{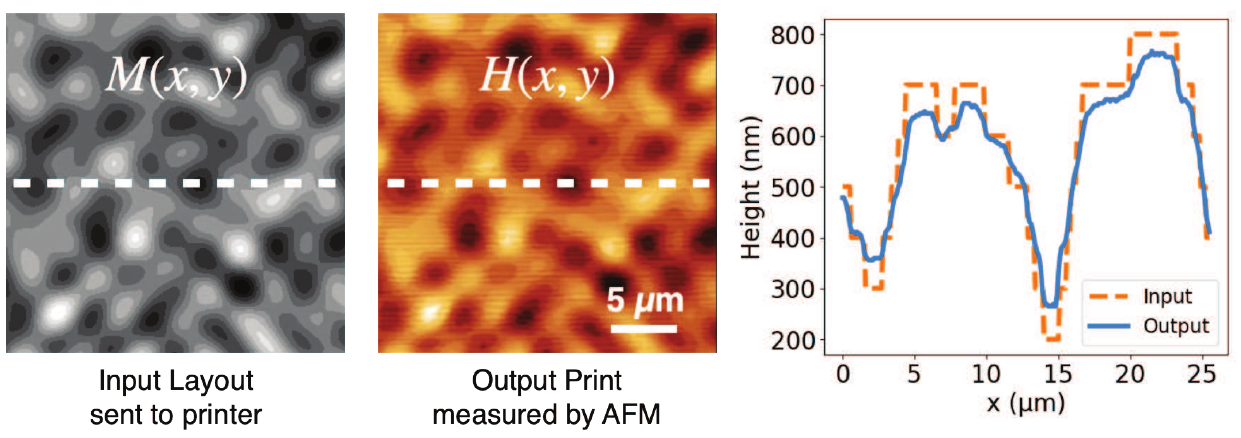}
\vspace{-3mm}
 \caption{An example of the experimental dataset for learning a neural lithography simulator. The line profiles along the white dashed lines show a clear height deviation between the input layout and the output print.}
\label{fig: dataset}
\end{figure}

To learn parameterized $\hat{g}_{\theta}$ that enables the optimization of structure, we collect the first $2.5\text{D}$ dataset $\mathcal{D} = \{(\mathbf{M}_i, \mathbf{H}_i)\}_{i=1}^{N}$ in photolithography, where $N=96$ is the size of the dataset, ${\mathbf{\mathbf{M}}_i}\in \mathbb{Z}_{\leq 12}^{n_1 \times n_2}$ is a random input layout and ${\mathbf{\mathbf{H}}_i}\in \mathbb{R}^{n_1 \times n_2}$ is the height profile of the fabricated structure, fabricated by a TPL system (Photonic Professional GT2, Nanoscribe GmbH), and measured offline by AFM. Here $n_1\times n_2 = 256 \times 256 $ is the lateral size of the structure. Since the slicing distance along the height direction is 0.1 $\mu m$, and we have $12$ levels in our fabrication, the input layouts represent discrete height values from $0$ to $1.2$ $\mu m$, providing a phase modulation range from $0$ to $2.07\pi$ given a refractive index difference $\Delta n = 0.548$ between the fabricated structure and air and illumination wavelength $\lambda=0.6328\mu m$. We register every AFM image $\mathbf{H}_i$ with its corresponding input layout ${\mathbf{M}_i}$ with the aid of homography estimation. Fig.~\ref{fig: dataset} shows an example data pair. 
Details on the fabrication and characterization of the prints are in the Supplements.

\subsection{End-to-end co-optimizing design and manufacturability in computational optics}~\label{subsec: downstream tasks}
\begin{figure*}[h!]
\centering
\includegraphics[width=.95\linewidth]{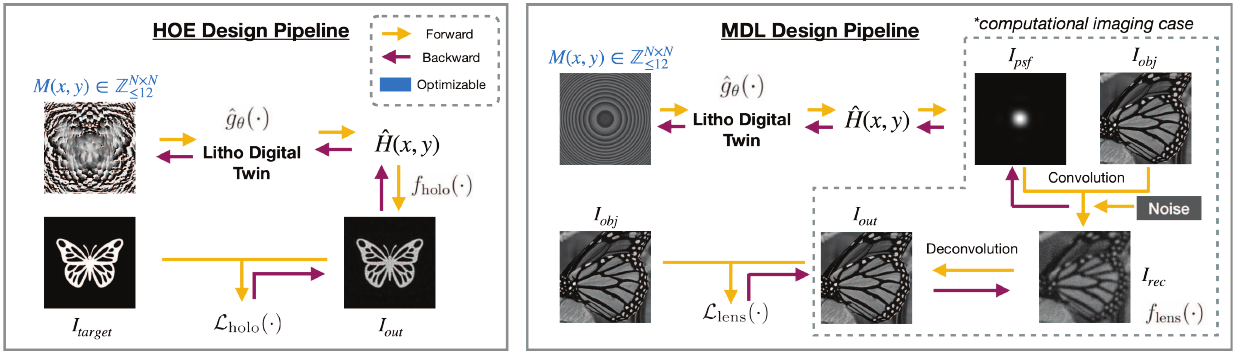}
\caption{\textbf{Design pipelines of holographic optical elements (HOE) and multi-level diffractive lens (MDL).} }
\label{fig: design pipeline}
\end{figure*}

We validate the performance of neural lithography with two representative tasks in computational optics. Below, we present the light transport and design pipeline for each task in Fig.~\ref{fig: design pipeline}. Due to the low absorption property of the fabricated optical element ~\cite{li2019uv}, we treat it as a thin phase object with phase profile $\phi\left(x,y\right) = \frac{2\pi\Delta n}{\lambda}{H}\left(x,y\right)$. Our work focuses on accurate $H(x,y)$ and omits the variation in $\Delta n$ as it was reported to be below $0.003$, the phase variation induced by which is much smaller than that caused by variation of $H(x,y)$~\cite{dottermusch2019exposure}. In the design, we use the output $\hat{H}(x,y)$ from our lithography digital twin $\hat{g}_{\theta}$ for $H(x,y)$ and optimize $M(x,y)$ via auto-differentiation through wave-optics-based forward light transport. We utilize the Gumbel-Softmax reparameterization trick~\cite{jang2016categorical} to differentiate through the discrete design layout ${{M}}\in \mathbb{Z}_{\leq12}^{n_1 \times n_2}$ (see details in Supplements).

\subsubsection{Holographic optical element}~\label{subsec: design of HOE}
A holographic optical element (HOE) is a micro-structured optical component that generates a desired image or diffraction orders and is commonly used in beam splitting, 3D imaging/display, and augmented reality. We design a HOE for an inline holography system, where the fabricated HOE is expected to generate a pre-designed holographic image at imaging distance $\Delta z = 300\mu m$ upon plane wave illumination. We substitute $f_{\text{optics}}$ in Eq. ~\ref{eq: obj computational optics} with $f_{\text{holo}}$ that calculates the intensity of the electric field $I_{\text{out}}$ after free-space wave propagation of distance $\Delta z$ with Rayleigh–Sommerfeld convolution~\cite{goodman2005introduction}:
\begin{subequations}

\begin{align}
    I_{\text{out}}(x,y) &= f_{\text{holo}}\left(\phi(x,y)\right), \\
    &= \left| e^{j\phi(x,y)}*h_{\Delta z}(x,y)\right|^2,
\end{align}
\end{subequations}
where $h_{\Delta z}(x,y) = \frac{\Delta z}{2\pi r^2}(\frac{1}{r}-\frac{j2\pi}{\lambda})e^{\frac{j2\pi r}{\lambda}}$ is the propagation kernel with $r=\sqrt{x^2+y^2+\Delta z^2}$. Given a target holographic image $I_{\text{target}}$, the loss function $\mathcal{L_{\text{holo}}}$ is:

\begin{align}
\mathcal{L_{\text{holo}}}(I_{\text{target}}, I_{\text{out}}) = \mathcal{L_{\text{rmse}}}(I_{\text{target}}, I_{\text{out}}) + 0.1 (1-\eta_{\text{eff}}(I_{\text{out}})),
\end{align}
where $\mathcal{L_{\text{rmse}}}$ is the root mean square error loss, $\eta_{\text{eff}}(I_{\text{out}})$ is the energy efficiency,  calculated as the sum of intensity inside the region of $I_{\text{target}}>0$ at image plane divided by the sum of total illumination light intensity. 

\subsubsection{Multi-level diffractive lens}~\label{subsec: design of lens}
Compared to conventional refractive lenses, diffractive lenses can perform imaging tasks while being lightweight and compact~\cite{banerji2019imaging}. We design MDLs with focal length $\Delta z_{f} = 400\mu m$ and object to lens distance is $\Delta z_{\text{ol}} =140 mm$. The system's point spread function (PSF) $I_{\text{psf}}$ is the intensity field illuminated by a point source at the object plane.  In the context of incoherent imaging under paraxial approximation, the image formation is a shift-invariant convolution of the object $I_{\text{obj}}$ with $I_{\text{psf}}$, and the camera records:

\begin{subequations}
\begin{align}
    I_{\text{rec}}(x,y) 
    &= \left| e^{j\phi(x,y)}*h_{\Delta z_{f}}(x,y)\right|^2*I_{\text{obj}}(x,y)+\eta,\\
    &= I_{\text{psf}}(x,y)*I_{\text{obj}}(x,y)+\eta,
\end{align}
\end{subequations}
where $\eta \sim \mathcal{N}(0, \sigma^2)$ is additive Gaussian readout noise.
The propagation from the object to the lens is ignored in the $I_{\text{psf}}$ calculation because $\Delta z_{\text{ol}}$ is much larger than the diameter of the lens ($120 \mu m$ in our case) and we can assume the wavefront from the point source to be flat. Here, we design two types of MDL. The first one is for direct imaging without post-processing the acquired camera recording $I_{\text{rec}}$, and design metric $\mathcal{L}_{\text{lens}}$ is the focusing efficiency~\cite{banerji2019imaging} of $I_{\text{psf}}$; the second one is for in-direct imaging (namely, computational imaging), where we assess the imaging quality after deconvolving $I_{\text{rec}}$ by $I_{\text{psf}}$.

\textbf{Direct imaging:} We calculate the PSF's center-to-background signal ratio (CBR) $r_{cbr}$ and the corresponding loss function $\mathcal{L_{\text{lens}}}(I_{\text{psf}}) = -\text{log}(r_{cbr})$ . Considering that we have a numerical aperture $NA\approx0.15$, corresponding to a diffraction-limited focal spot size of $ \approx4\mu m$, we choose center region size to be $1 \mu m \times 1 \mu m$ and we empirically find this setting leads to the sharpest PSF after the optimization.

y\textbf{In-direct/computational imaging:} We evaluate the image quality after post-processing~\cite{sitzmann2018end,tseng2021neural}. we apply a fixed Richard-Lucy deconvolution process and calculate the mean absolute error (MAE) between the ground truth $I_{\text{obj}}$ and deconvolved result as the loss $\mathcal{L_\text{lens}}$:
\begin{equation}
    \mathcal{L}_{\text{lens}}(I_{\text{rec}}, I_{\text{obj}}) = \mathcal{L}_{\text{mae}}(f_{\text{deconv}}(I_{\text{rec}}, I_{\text{psf}}), I_{\text{obj}}),
\end{equation}
where $f_{\text{deconv}}$ denotes the deconvolution process, and in this work, we use the Richard-Lucy deconvolution algorithm~\cite{zhao2018nonlinear} with $50$ iterations.

\section{Results}
We first evaluate the accuracy and generalizability of our neural lithography simulator in subsection~\ref{subsec: res, diff litho simulator} (lower-level optimization)\textcolor{red}{}. Then we present the results of designed and fabricated optical elements in computational optics tasks, in subsections~\ref{subsec: result of HOE} and~\ref{subsec: result of lens} (upper-level optimization), which show the priority of our pipeline in 
mitigating the design-to-manufacturing gap. We further provide detailed results to visualize the improvement by our pipeline in subsection~\ref{subsec: d2m} and noise analysis to investigate the performance limit of our method in subsection~\ref{subsec: noise analysis}.

\subsection{Forward predicting capability}\label{subsec: res, diff litho simulator}

\begin{figure}[h!]
\centering
\includegraphics[width=0.95\linewidth]{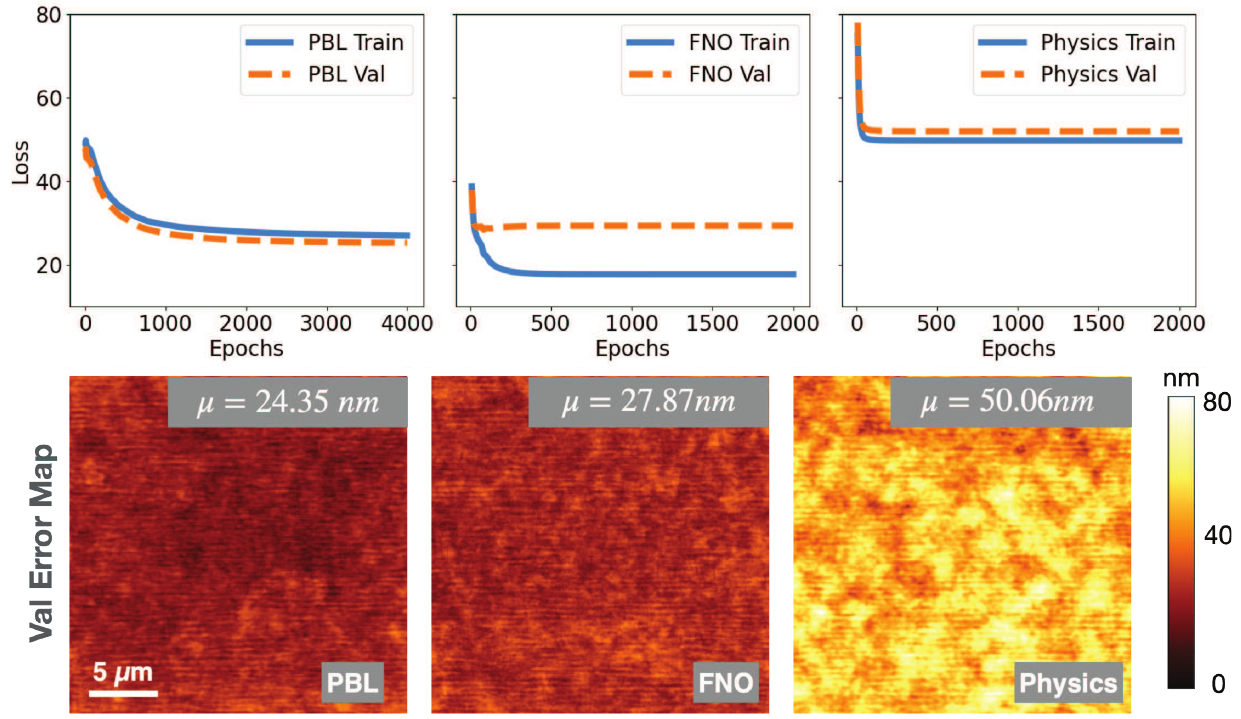}
\caption{\textbf{Forward predicting capability of the $\hat{g}_{\theta}$.} Top: The training and validation loss curves correspond to the three models explored in our work. Bottom: The corresponding average validation error map and the mean error value across the map.}
\label{fig: model capacity}
\end{figure}

We evaluate the efficacy for $\hat{g}_{\theta}$, trained from the dataset described in section~\ref{subsec: dataset} and Fig~\ref{fig: dataset}. We visualize the loss curves and error maps that compare our PBL model with other learnable modeling approaches, including the parameterized physics model~\cite{lang2022towards} and Fourier Neural Operator (FNO)~\cite{yang2022generic}. The details of the models and training hyperparameters are in the Supplements. We show that the PBL modeling receives the lowest validation loss and error. The parameterized physics model has limited degrees of freedom and thus cannot fully fit the data. Without applying the physics model as a prior, FNO exhibits good fitting ability while tending to overfit on the training set.  

\subsection{Application: Holographic optical element}~\label{subsec: result of HOE}
\begin{figure}[h!]
\centering
\includegraphics[width=.95\linewidth]{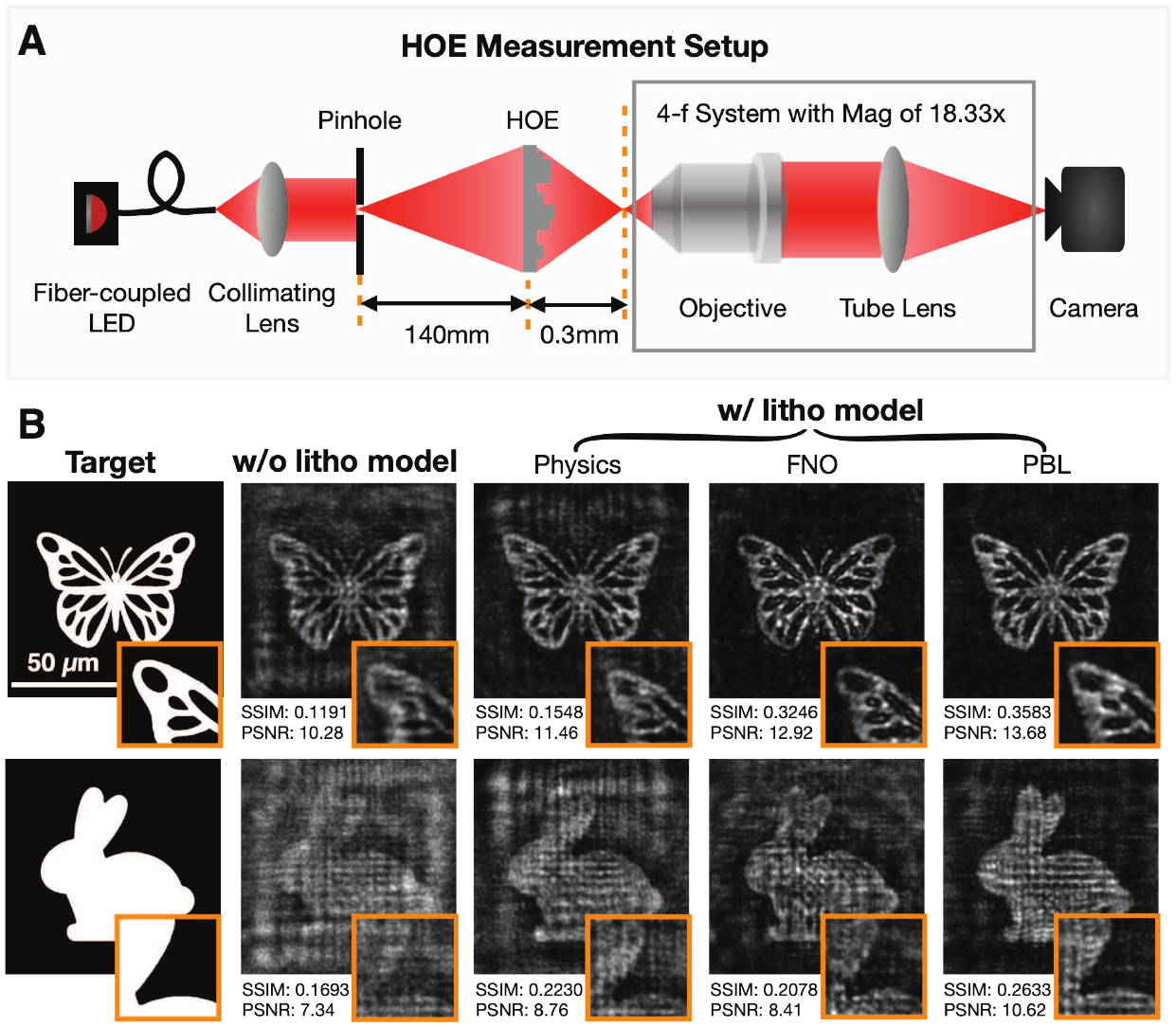}
\vspace{-2mm}
\caption{\textbf{Performance of the designed HOE.} A: Sketch of the setup for characterizing the performance of HOE. B: Performance comparison shows that with the lithography model, the quality of the holographic images is improved, and the PBL model brings the most pronounced improvement.}
\label{fig: holo setup}
\end{figure}
We design several HOEs with the three lithography models using the pipeline in subsection ~\ref{subsec: design of HOE}. The traditional design without a lithography model in the loop is also added as a reference. The designed input layouts are $102.4\mu m \times 102.4\mu m$ with an optimizable pitch size of $0.4 \mu m$ and optimized with Adam for 1000 iterations under a learning rate of $0.1$. We capture holographic images of the fabricated HOEs with the setup sketched in Fig.~\ref{fig: holo setup}A. 
As expected, the resulting images with lithography models present better image quality and contrast. We also assess the SSIM and PSNR of the images using the Kornia package~\cite{riba2020kornia}, where the one optimized using the PBL model scores the highest.
The overall low value of the SSIM and PSNR may come from the measurement system's misalignment and the layout's limited quantization levels.

\subsection{Application: Multi-level diffractive lens}~\label{subsec: result of lens}
\begin{figure}[h!]
\centering
\includegraphics[width=.95\linewidth]{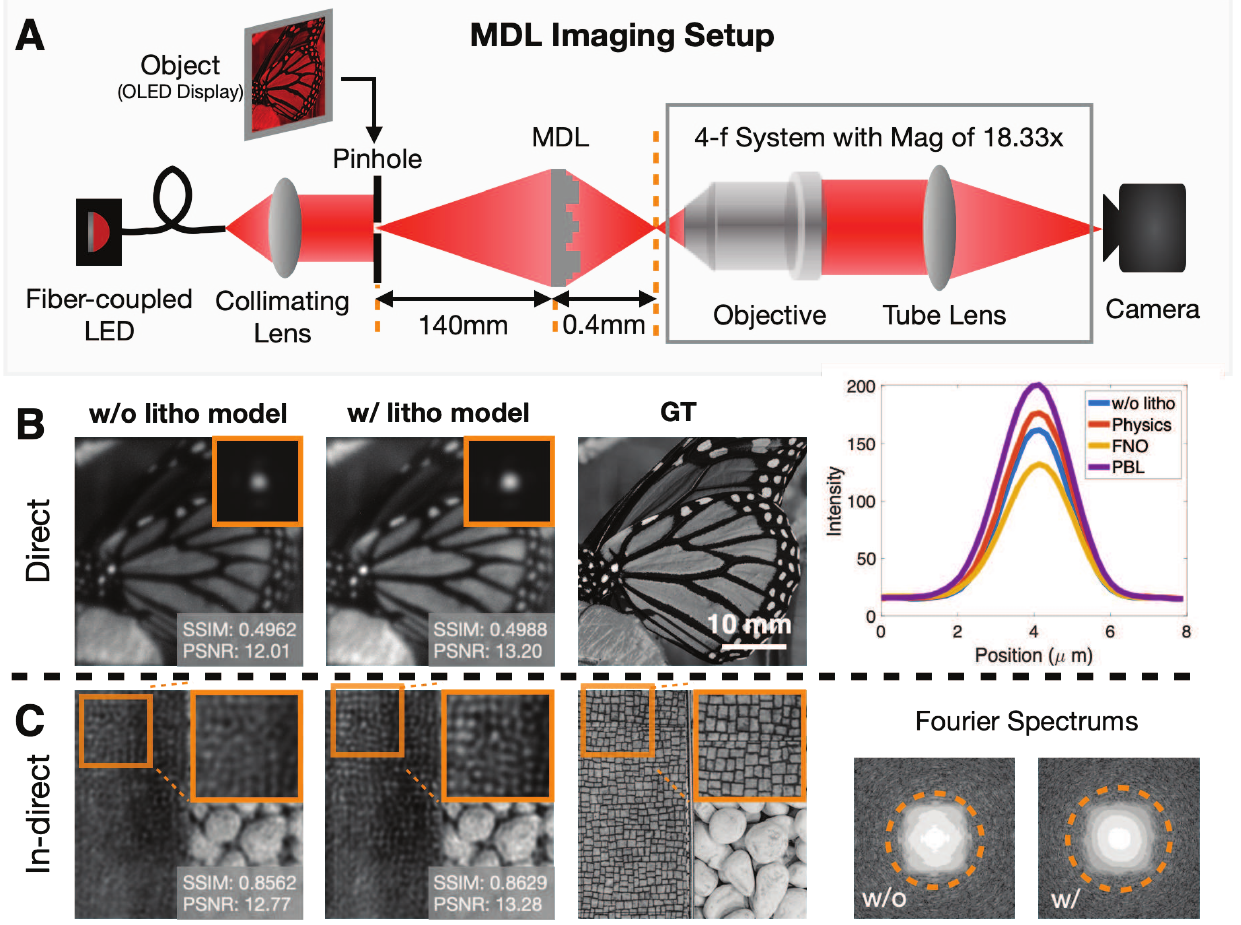}
\vspace{-2mm}
\caption{\textbf{Imaging performance with the designed MDL.} A: Sketch of the setup for characterizing the performance of MDL. B: We show our measured PSFs and direct imaging results (i.e., w/o deconvolution) corresponding to design w/o and w/ PBL litho model. The end of this row shows the line profiles of PSFs designed w/o or w/ different litho models. C: Computational/Indirect Imaging result of the MDL. The lower right compares the Fourier spectrum of the designed PSFs. Our method's design enhances the contrast in direct imaging (B) and the high-frequency imaging performance in computational imaging (C).
}
\label{fig: performance of MDL}
\end{figure}

We compare the performance of our designed and fabricated MDLs to investigate the performance gain brought by the neural lithography (Fig.~\ref{fig: performance of MDL}). The designed MDL layouts are $120\mu m \times 120\mu m$ with an optimizable pitch size of $0.1 \mu m$ and optimized with AdamW for 1000 iterations under a learning rate of $3$. We built a system sketched in Fig.~\ref{fig: performance of MDL}A, where the pinhole is used to calibrate the PSF of the imaging system. During imaging, we switch the light path to the object plane (the OLED display), which is in conjugation with the pinhole plane.

The \textbf{direct imaging} outcomes (Fig.~\ref{fig: performance of MDL}B) demonstrate the imaging contrast bolstered by the PBL lithography simulator's inclusion in the design loop. Further profiling of PSFs associated with various MDLs reveals that the design aided with the PBL model yields the brightest PSF. Furthermore, the \textbf{computational imaging} result in Fig.~\ref{fig: performance of MDL}C shows that when both deconvolving with the camera-captured (calibrated) PSFs, the one with the neural lithography simulator results in more details, which retains the higher-frequency imaging capability, as highlighted in the zoom-in view. Post-calibration has been used to mitigate fabrication inaccuracy in previous work~\cite{tseng2021neural}, but the comparison here shows that it cannot fully address the issue. This is also evidenced by the Fourier spectrum of the PSFs shown on the right-hand side of Fig.~\ref{fig: performance of MDL}C, where the Fourier spectrum corresponding to the PSF associated with the PBL litho model exhibits higher frequency coverage. 

\begin{figure}
\centering
\includegraphics[width=.9\linewidth]{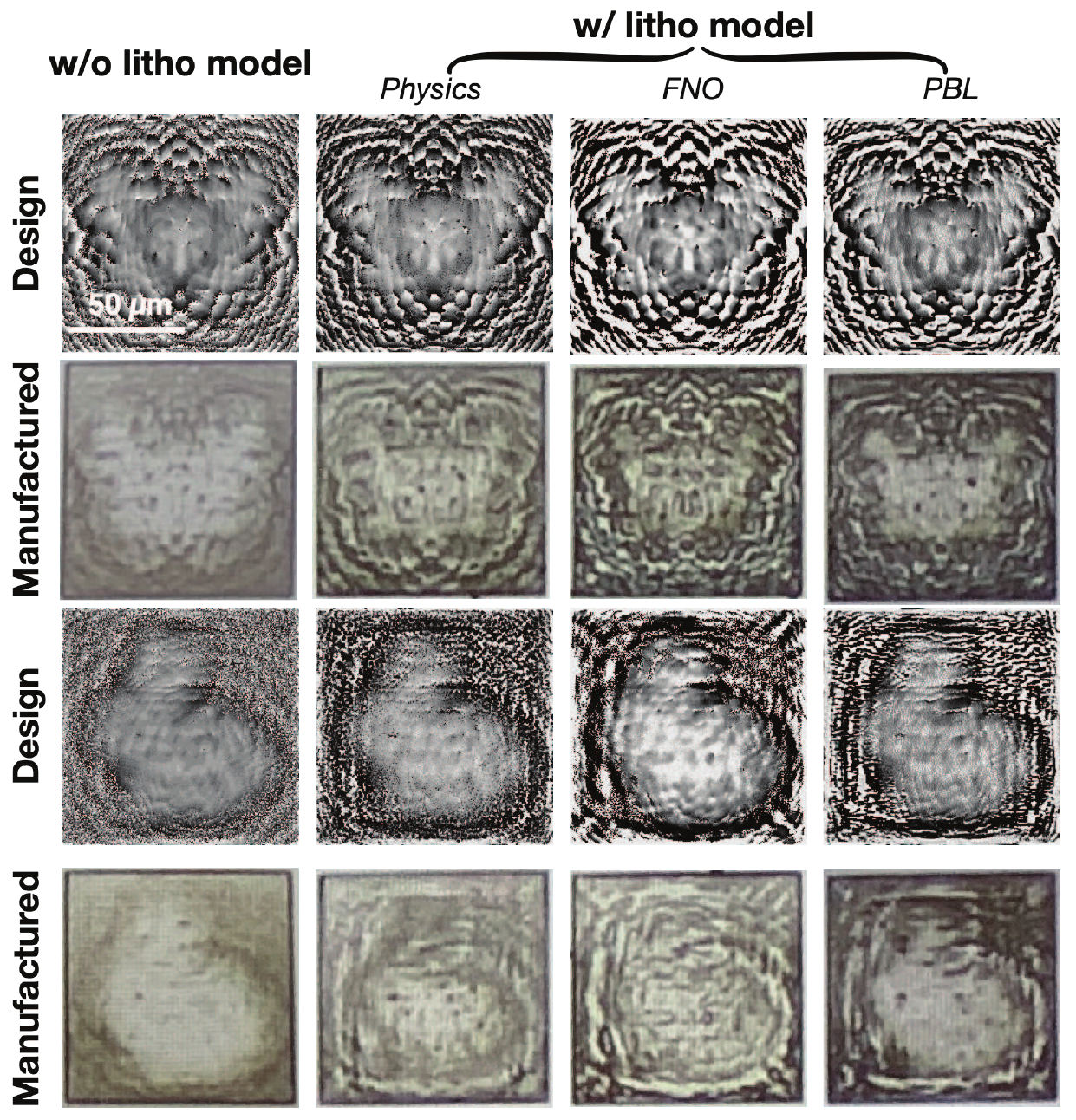}
\caption{\textbf{Qualitative comparison on the images of designed HOE and the corresponding fabricated structures.} The designed structure from one w/ litho model shares more similarity with their fabricated correspondences. Designs w/o the litho model render too fine-detail layout mask, the fabricates of which deviate a lot from the original designs. The slight difference in optical micrograph color tones comes from non-uniform illumination.}
\label{fig: HOE design and fabrcaites}
\end{figure}

\begin{figure}
\centering
\includegraphics[width=.9\linewidth]{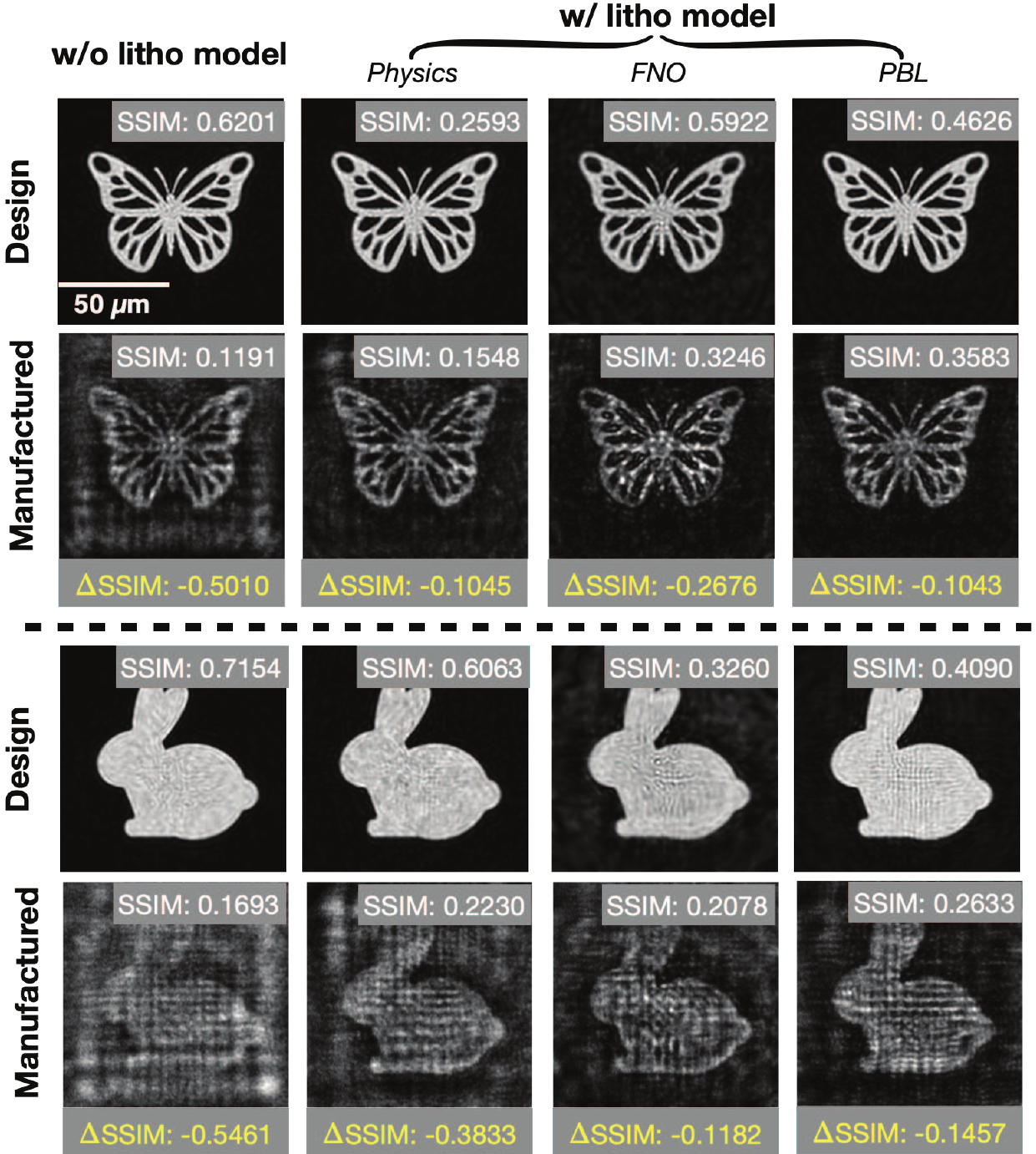}
\caption{\textbf{Quantitative comparison of holographic images generated by HOE in the design and real experiment.} $\Delta \text{SSIM}$ between the holographic images of the designed and that of manufactured HOEs shows that a huge design-to-manufacturing gap exists for the naive design w/o considering the lithography process. Though not exhibiting good SSIM in the design, the printed HOEs w/ the PBL model result in the best SSIM.}
\label{fig: HOE results}
\end{figure}

\begin{figure}
\centering
\includegraphics[width=.9\linewidth]{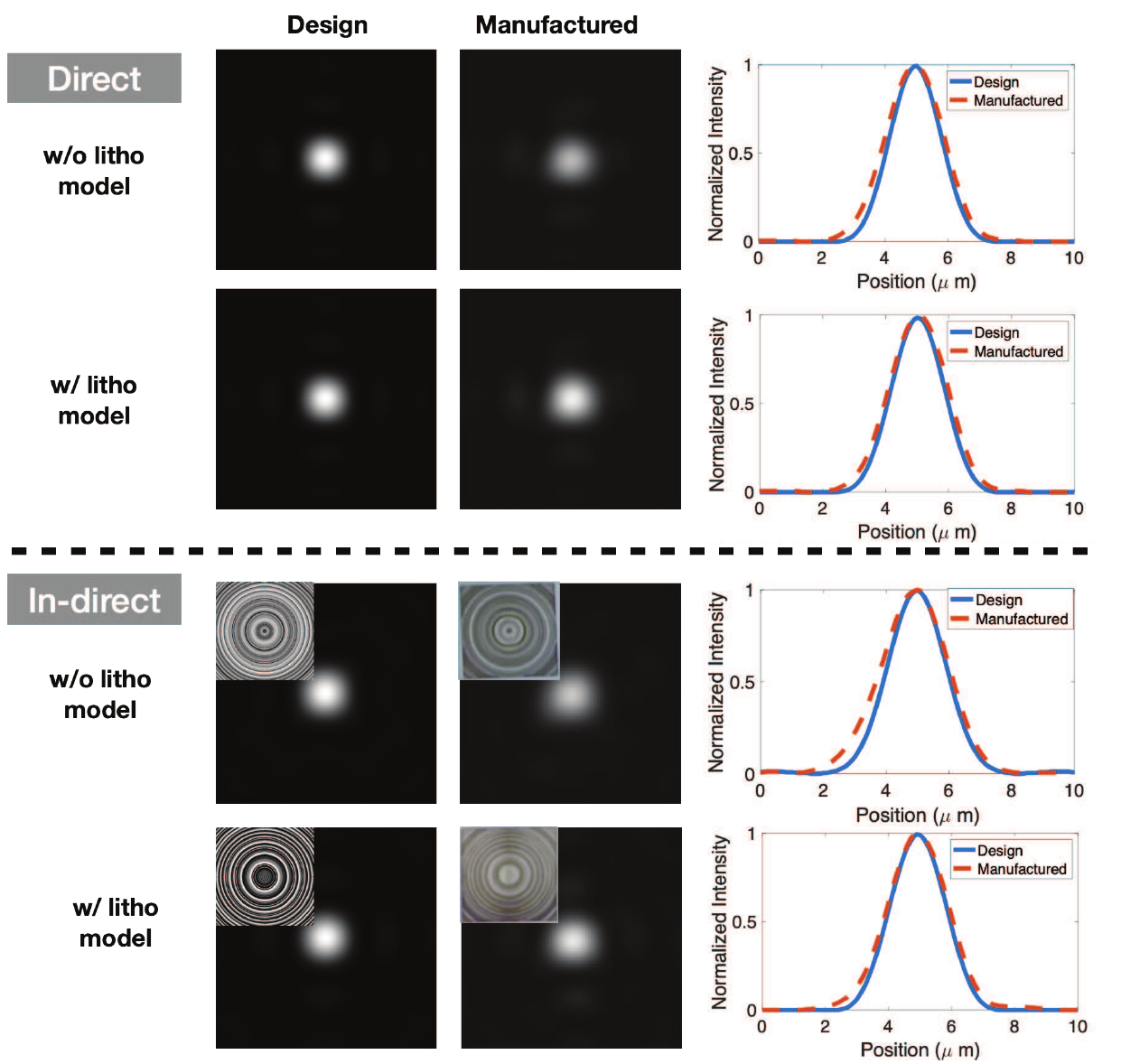}
\caption{\textbf{Comparison of PSFs generated by MDLs in the design and real experiment.} In both the direct and indirect/computational imaging setting, the naive design w/o lithography model has a larger deviation between the shape from the designed and experimental PSF. In contrast, the deviation is small when we apply neural lithography.}
\label{fig: PSF profiles}
\end{figure}

\subsection{Design-to-manufacturing gap mitigation}\label{subsec: d2m}

In Fig.~\ref{fig: HOE design and fabrcaites}, we show the layout of the designed HOEs and optical micrographs of the corresponding fabricated structure. We observe more high-frequency features in the designs without the lithography model and a smoother profile in the designs with the lithography model. 
Also, there is a better coincidence between designs and prints when using our PBL lithography model in the design loop. In contrast, a large mismatch exists between design and print without a neural photolithography simulator. This evidences the lithography model serves as a regularizer on fabrication feasibility during the design process.

A quantitative comparison of the holographic images of the HOEs (micrographs of which are in Fig.~\ref{fig: HOE design and fabrcaites}) in both design and experiment in Fig.~\ref{fig: HOE results} reinforces our conclusion. The ones without the lithography model show the highest SSIM in the design while being the worst in the experimental measurement. Given the huge design-to-manufacturing gap in Fig.~\ref{fig: HOE results}, it is unsurprising. The images using our neural photolithography simulator show the highest SSIM in manufacturing while maintaining the lowest gap between the design and manufacturing values. The comparison between the \textit{Physics} and \textit{PBL} models in the figure reveals that the \textit{PBL} model does more than just add parameterized smoothness to the design profile, as the \textit{Physics} model does. Instead, it captures more complex physical relationships, thereby better mitigating the design-to-manufacturing gap.

Similarly, the effectiveness of our method in creating multi-level diffractive lenses is evidenced by the PSFs displayed in Fig.~\ref{fig: PSF profiles}. In both direct and computational imaging scenarios, the line profiles across the center of the PSFs from MDLs designed using the lithography model show a better fit to the design than those lacking the lithography model. Additionally, in the direct imaging case, the PSF with the lithography model displays higher overall brightness. In the computational imaging case, there is greater correspondence between the design and the resulting optical micrograph when using the lithography model.

\subsection{Analyze noise to the aleatoric uncertainty}\label{subsec: noise analysis}

\begin{figure}[h]
\centering
\includegraphics[width=0.9\linewidth]{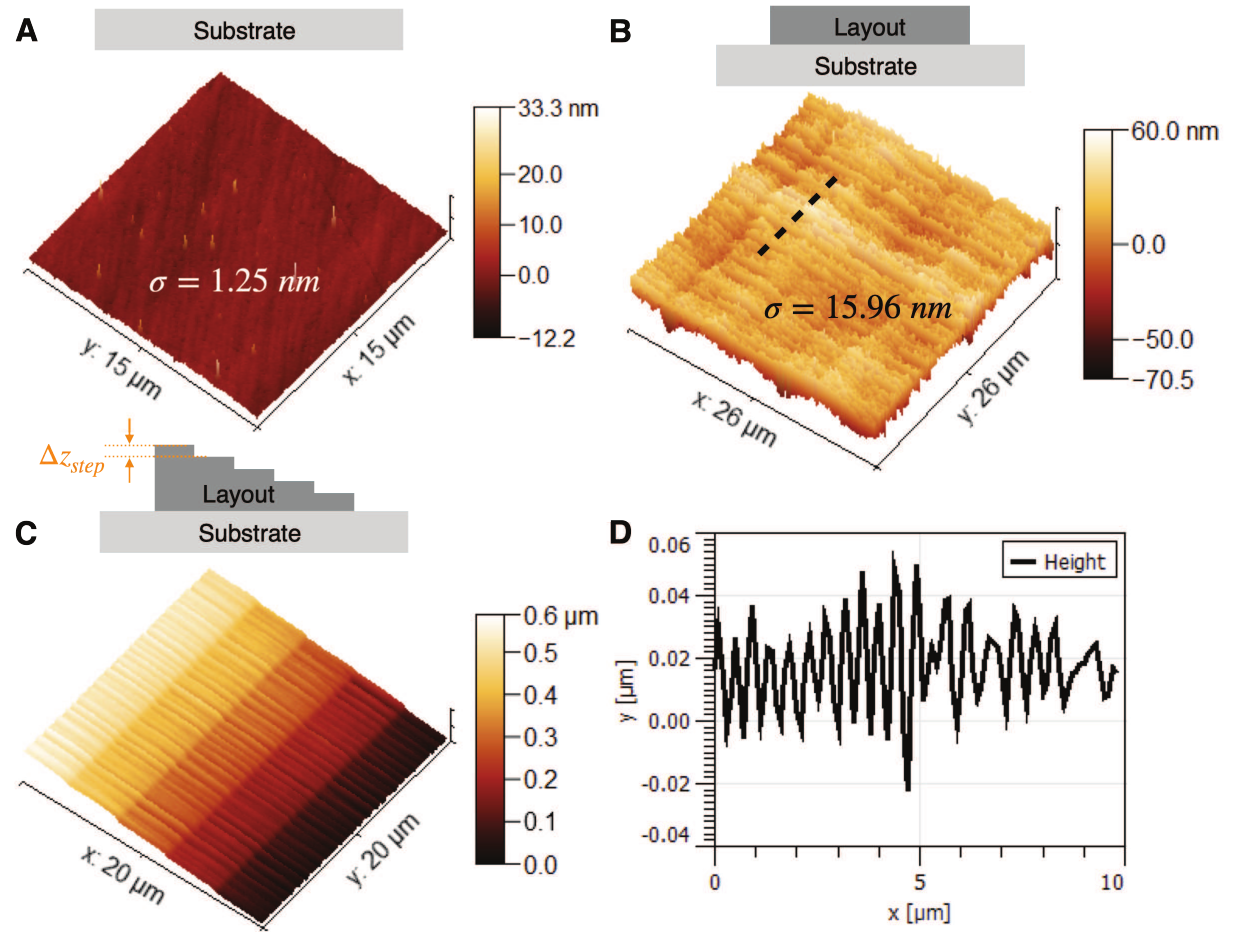}
\caption{\textbf{Aleatoric uncertainty limits the upper bound of our pipeline's performance.} We measure a clean substrate (A), a flat print (B), and a 5-step print (C) sitting on top of the substrates to visualize and quantify the data noise. The line profile along the black dashed line in B is further visualized in D. The 5-step print with step interval $\Delta z_{\text{step}} = 0.1\mu m$ in C shows that the data uniformity is high at a larger scale, while the measured flat surface in B indicates there is a certain degree of uncontrolled noise at the nanoscale from fabrication and measurements. The comparison between A (w/o prints) and B (w/ prints) shows the prints contribute to the most noise. The line profile in D shows strong randomness in the fabrication error, which makes it hard to model. 
} 
\label{fig: aleatoric uncertainty}
\end{figure}

The ultimate predicting ability of our lithography simulator depends on both the model's learning ability and the dataset's quality, relating to the model's epistemic uncertainty and the dataset's aleatoric uncertainty, respectively~\cite{hullermeier2021aleatoric}. In Fig.~\ref{fig: aleatoric uncertainty}, we visualize and quantify noises by fabricating a flat surface (Fig.~\ref{fig: aleatoric uncertainty}B) and a 5-step structure (Fig.~\ref{fig: aleatoric uncertainty}C) and measuring them together with a flat substrate (Fig.~\ref{fig: aleatoric uncertainty}A). 
 \textit{Comparison between Fig.~\ref{fig: aleatoric uncertainty}A and Fig.~\ref{fig: aleatoric uncertainty}B shows that the unmodeled fabrication error is the dominant noise source in our TPL demonstration.} While the roughness of a clean substrate without printed features is $\sigma = 1.25 nm$ (Fig.~\ref{fig: aleatoric uncertainty}A), the printed flat surface shows roughness of $\sigma = 15.96 nm$, resulting from the non-periodic line-shaped features post-fabrication.
  We further plot a height profile of these features along the black dashed line on the flat surface in Fig.~\ref{fig: aleatoric uncertainty}D, where the randomness in the plot evidences these line-shaped features are hard to model. We also observe these features in the fabricated 5-step structure, but the $0.1 \mu m$ step size can still be well identified, indicating the structure is well recognized in large structures. Given these, our predicting error of $\mu = 24.35 nm$ using PBL indicates future room to improve our $\hat{g}_{\theta}$, such as with more training data or more intricate model considering these line-shaped features, or even switching to projective photolithography systems. Nevertheless, $\sigma = 15.96 nm$ reveals an upper performance bound of current $\hat{g}_{\theta}$ given the TPL system we use. 

\section{Discussion}

\textbf{Limitations:}
\begin{itemize}
    \item The data noise during fabrication and measurement fundamentally limits the optimization capability. 
    \item The neural photolithography simulator has no theoretical guarantees, which might produce adverse designs in the ill-posed inverse design procedures.
    \item The sim2real gap between simulated and real optical systems impairs performance. The post-calibration methods mitigate the sim2real gap but don't correct the bias from incorrect $f_{optics}$ modeling during design. 
\end{itemize}

\noindent \textbf{Future work:}
As a first work on differentiable/neural lithography, our work is poised to inspire future investigations:

\textit{Applying to other design and fabrication processes.} Our real2sim pipeline is adaptable to various lithography techniques, each requiring fine-tuning the model architecture. For example, extreme ultraviolet (EUV) lithography necessitates extra modeling for reflective projection optics~\cite{wagner2010lithography}, while etching-based or nanoimprint methods require an added step to transform the resist profile into the final structure~\cite{barcelo2016nanoimprint}. Similarly, e-beam lithography involves a more complex PSF from electron scattering~\cite{koleva2018modeling}. Moreover, future work can investigate applying our end-to-end co-optimization pipeline to design and fabricate more complex downstream tasks, such as obtaining depth profile from defocus~\cite{ikoma2021depth} and implementing optical computing~\cite{goi2022direct}.  

\textit{Further improving the pipeline.}
On the hardware side, other characterization tools, such as 3D SEM~\cite{tafti2015recent} or EUV diffractive imaging~\cite{gardner2017subwavelength}, could be further explored for collecting high-throughput and high-accuracy datasets. On the algorithm side, comprehensive 3D modeling instead of 2.5D will cover more complex resist profiles. We can also further incorporate advanced modeling like neural architecture searching~\cite{ba2019blending} to improve the precision of photolithography predictions or utilize implicit neural fields~\cite{yang2021geometry} to augment the design efficiency of neural lithography.

\noindent \textbf{Conclusion:}
To conclude, our study unveils a unique perspective of jointly optimizing diffractive optical elements in computational optics and their fabrication feasibility. By employing a 'real2sim' learned digital twin of the photolithography system, we enhance the outcomes of optical tasks. This methodology paves the way for the precise manufacturing of sophisticated computational optics elements, expediting their introduction to commercial products. We anticipate that the 'real2sim' approach presented in this paper will encourage both industry and academia to complement traditional white-box methods with the more adaptive, data-driven gray-box methodologies for modeling computational optics systems, as exemplified by the lithography system discussed in our work.

\begin{acks}

This work is supported by NIH (5-P41-EB015871), Fujikura Limited, and the Hong Kong Innovation and Technology Fund (ITS/178/20FP). This work was carried out in part using MIT.nano's facilities. 

\end{acks}

\bibliographystyle{ACM-Reference-Format}
\bibliography{bibliography}

\appendix

\addcontentsline{toc}{section}{Supplements}

\section{Differentiating through multi-level diffractive optical elements with Gumbel-Softmax trick} 
The leader optimization in the main text requires the pipeline to be fully differentiable. However, the quantization levels of the mask layout are limited in many photolithography fabrication processes, including the two-photon lithography (TPL) system we use in this work. The Nanoscribe TPL system has a minimum axial slicing distance of $0.1~\mu m$. That is to say, it only supports discrete height inputs instead of a continuous height profile between 0 and $2 \pi$, and we calculate that it has 12 discrete levels, leading to ${M}\in \mathbb{Z}_{\leq12}^{n_1 \times n_2}$, to be random discrete values from $\mathbb{Z}_{\leq12} = \{0,1, ..., 12\}$\footnote{Here we abuse the use of notations. Normally $\mathbb{Z}$ contains $\{..., -2, -1, 0, 1, 2, ...\}$ and here we refer it as $\{0, 1, 2, ...\}$.}. Thus, we apply the Gumbel-Softmax trick~\cite{jang2016categorical} on the categorical height variables to allow for sampling from the categorical distribution during the forward pass through our lithography simulator. It is worth mentioning that this trick has recently been used in other computational optics tasks ~\cite{choi2022time, zhu2022infrared}.     

For each pixel $r_{pq}$ in the layout $M$, we model it follows categorical distribution with class probabilities $\pi_0, \pi_1, ...\pi_{12}$, corresponding to the $i = 0, 1, ..., 12$ categories, i.e., quantized phase levels. 

We introduce Gumbel noise to add randomness to the sampling procedure.
\begin{equation}
    g_i = -\log (-log(u_i)),\;u_i \sim \text{Uniform}(0,1),
\end{equation}
where  $g_i$ is the Gumbel noise. We then calculate the vector $y_i$ using softmax as a differentiable approximation to the argmax-based sampling:
\begin{equation}
   y_i = \text{Softmax}[(g_i + \log \pi_i)/\tau], 
\end{equation}  
where we choose temperature $\tau=0.5$ in our experiment. See more explanations of Gumble-Softmax in~\cite{jang2016categorical}.

\begin{figure*}[ht]
\centering
\includegraphics[width=.8\linewidth]{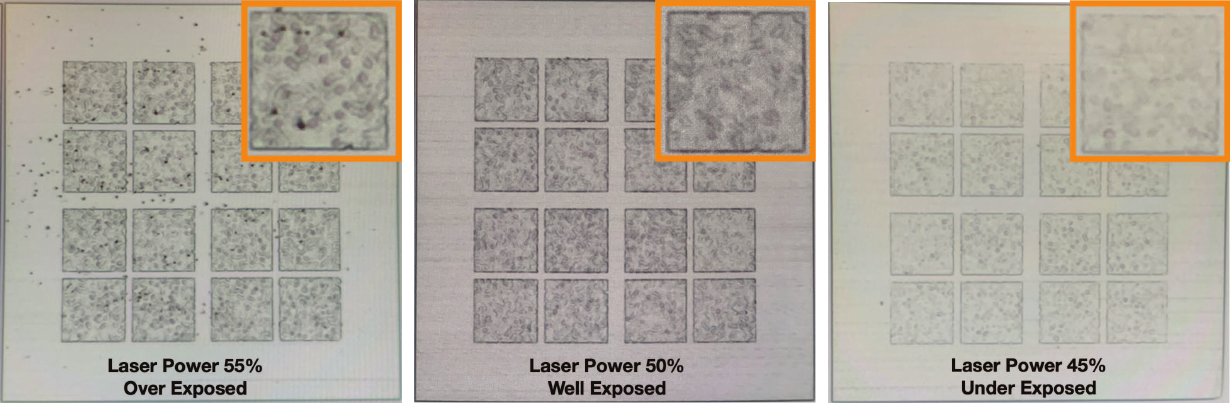}
\caption{\textbf{$50\%$ laser power is suitable for the lithography system we use.} }
\label{fig: laser power}
\end{figure*}

\section{Fabrication and characterization}~\label{subsec: fab and afm details}
\subsection{Mask and optical elements fabrication}
We fabricate the masks in dataset $\mathcal{D}$ and computational optical elements in downstream tasks using a commercial two-photon lithography system -- Photonic Professional GT2 (Nanoscribe GmbH). We use a 63×/1.4NA objective, IP-Dip photoresist, and a $25 mm \times 25 mm \times 0.7 mm$ fused silica glass substrate. 

We follow the printing parameter set according to ~\cite{wang2020toward}, reproducing very well in our task. 
The writing parameters for the photolithography process are fixed for all prints: the ratio of laser power $50\%$, scan speed $40000$ $\mu m/s$, and hatching and slicing distance of 0.1 $\mu m$. 

The laser power ratio is critical for the lithography performance. We grid-search the laser power in case it leads to poor manufacturing. As seen from Fig.~\ref{fig: laser power}, a laser power $50\%$ is the most suitable for printing the structure and would neither over- nor under-expose the photoresist. 

A 1-$\mu m$-thick base layer is printed with the same parameters before printing the structures to compensate for errors in identifying the position of the substrate and resist interface. After photopolymerization, the structures are developed in resist developer polyethylene glycol methyl ether acetate (PGMEA) for 20 min, followed by immersion in isopropyl alcohol for 5 min. Then, we use a nitrogen gun to dry the structures before deployment.

\subsection{Prints characterization}
We measure the height profiles of the fabricated structures using Jupiter XR atomic force microscope (AFM) (Asylum Research, Oxford Instruments). We choose to use the AFM based on the performance comparison of the three mainstream methods: the Profilometer~\cite{hong2021three}, SEM~\cite{liao2022line}, and AFM~\cite{chevalier2021rigorous}, which all have been used for characterizing the nano-optical prints before. AFM provides quantitative information on structure height with a high lateral resolution, while SEM doesn't have quantitative height information, and Profilometer doesn't have enough lateral resolution. 

\begin{table}[ht]
\fontsize{8pt}{8pt}\selectfont
    \centering
    \begin{tabular}{c|p{1.6cm}|p{1.5cm}|p{1.5cm}}
    \toprule
    &\textbf{Profilometer} & \hfil\textbf{SEM} & \hfil\textbf {AFM}  \\\midrule
    Lateral Resolution & \hfil\cellcolor{red!20} Low & \hfil\cellcolor{green!20}High & \hfil\cellcolor{green!20}High \\
    Speed & \hfil\cellcolor{green!20} Fast & \hfil\cellcolor{yellow!20} Medium &  \hfil\cellcolor{red!20} Slow \\
    Dimension & \hfil\cellcolor{green!20} 2.5D & \hfil\cellcolor{red!20} 2D & \hfil\cellcolor{green!20} 2.5D  \\
    \bottomrule
    \end{tabular}
    \caption{\textbf{Comparison of characterization methods used in the nano-optics.} Though slow in acquisition speed, AFM offers the required $2.5$D and high-lateral resolution imaging capability and thus becomes the best option for this work.}
    \label{tab: comparison of characterization techniques}
\end{table}

We acquire the AFM images with a $0.1 \mu m$ scanning pixel size. We then process the images in Gwyddion\footnote{http://gwyddion.net/} software to remove the background with $2^{nd}$ order polynomial fit and shift the minimum height value to $0$.

\section{Generate randomized printing layout}
To efficiently explore the design space of layout $M$ to train a $\hat{g}_{\theta}$ for the inverse design, we design the height layout ${\mathbf{M}}_i\in \mathbb{Z}_{\leq12}^{n_1 \times n_2}$ ($i\in \{1, 2, 3, ..., 96\}$) to be random discrete values from $\mathbb{Z}_{\leq12} = \{0,1, ..., 12\}$. The mathematical operation to synthesize $\mathbf{M}_i$ is in Eq.~\ref{eq: mask generate}, modified from~\cite{gostimirovic2022deep}. We sample a matrix $\mathbf{R}_{i}\in \mathbb{R}^{n_1 \times n_2}$ with each entry $r_{pq} \sim U(0,1)$.
We then apply a low-pass filter (LPF) on the Fourier transform $\mathcal{F}$ of this random matrix $\mathbf{R}_{i}$ to remove the high-frequency structural components; an inverse Fourier transform $\mathcal{F}^{-1}$ is applied on the normalized real part of it. Finally, we apply a quantization step $\mathcal{Q}$, which quantizes the layout pixel values to $\mathbb{Z}_{\leq 12}$.

\begin{subequations}\label{eq: mask generate}
\begin{equation}
    \mathbf{M}_i = \mathcal{Q}(norm\left\{ \Re\left\{\mathcal{F}^{-1}\left\{\mathcal{F}\left\{\mathbf{R}_i\right\}\cdot\text{LPF}\right\}\right\}\right\}),
\end{equation}

\begin{equation}
    \mathcal{Q}(\cdot) = \Delta \times \lfloor \frac{\cdot}{\Delta} + \frac{1}{2}\rfloor,
\end{equation}
\end{subequations}

where $\Delta=\frac{1}{12}$ is the step size for the quantization, $norm$ denotes normalizing the input to $[0,1]$. The final layouts contain structures of lateral sizes ranging from $0.1~\mu m$ to several microns.

\section{Model architecture and training settings}

\subsection{Training details}
When training the forward model $\hat{g}_{\theta}$ in the main text, we split the dataset $\mathcal{D}$ to train set size $72$ and validation set size $24$. We compare the three models: the PBL, the Fourier Neural Operator (FNO), and the parameterized physics model. These three models are trained using the Adam optimizer~\cite{kingma2014adam} and a batchsize of $4$ for $4000$ epochs. Specifically, the learning rate for training the PBL model is $0.0005$. 
Note that the neural lithography simulator should be retrained for every new fabrication system or a new photoresist/protocol.

\subsection{Model architecture}

\subsubsection{More details on the PBL model.}
{Table ~\ref{tab: params} summarizes the learnable parameters $\theta$ in lithography simulator $\hat{g}_{\theta}$}. There are single-value parameters in the corresponding Gaussian kernels and learnable weights and biases of the convolution layers inside each neural operator.

We do not include $\sigma_{\text{optical}}$ into $\theta$ as it is deterministic concerning the optical setup. $\sigma_{\text{optical}}$ is fixed as $200~nm$, calculated according to the beam focused by the illumination objective lens ~\cite{dong2003characterizing}. $\sigma_{\text{diffusion}}$ is learnable. To limit $\sigma_{\text{diffusion}}$'s range during the training, we further model $\sigma_{\text{diffusion}}$ as: 
\begin{equation}
    \sigma_{\text{diffusion}} = \frac{\eta}{1+e^{-\tau}},
\end{equation}
where $\eta$ is a hyper-parameter to limit the value range, and $\tau$ is the underlying learnable parameter. Specifically, in this work, we determine $\sigma_{\text{diffusion}}$'s value range to be [0, $2.5~\mu m$] according to~\cite{lang2022towards} and thus set the hyper-parameter
$\eta = 2.5~\mu m$.

\begin{table}[ht]
\fontsize{8pt}{8pt}\selectfont
    \centering
    \begin{tabular}{c|c}
    \toprule
    \textbf{Function} &\textbf{Parameter} \\
    \midrule
     $g_\text{thresh}$ & Weights and biases  \\
    $h_{\text{diffusion}}$ & $\tau$ \\
     $g_\text{shrink}$ & Weights and biases  \\
    $g_\text{mismatch}$ & Weights and biases \\
    \bottomrule
    \end{tabular}
    \caption{\textbf{Learnable parameters included in $\theta$ in PBL.}}
    \label{tab: params}
\end{table}

We apply the pointwise and areawise neural networks inside the physics-based neural network to model the corresponding operations inspired by the work from~\cite{tseng2022neural}. 

\textit{Pointwise neural operator} is used in $g_\text{thresh}$ and $g_\text{shrink}$. It affects the input at a pixel-wise level without affecting neighboring pixels. As summarized in Table ~\ref{tab: pointwise} below, it consists of three convolution layers using a $1 \times 1$ kernel, each followed by a Leaky Relu activation function (with the hyper-parameter negative slope $\alpha = 0.02$) except for the last layer.

\begin{table}[ht]
\fontsize{8pt}{8pt}\selectfont
    \centering
    \begin{tabular}{c|c|c|c}
    \toprule
    \textbf{Layer Name} &\textbf{Output Channels} &\textbf{Kernel Size} & \hfil\textbf {Activation} \\\midrule
    Conv1 & 128 & $1 \times 1$ & LeakyReLU \\
    Conv2 & 64 & $1 \times 1$ & LeakyReLU \\
    Conv3 & 1 & $1 \times 1$ & - \\
    \bottomrule
    \end{tabular}
    \caption{\textbf{Neural pointwise operator architecture.}}
    \label{tab: pointwise}
\end{table}

\textit{Areawise neural operator} is used in $g_\text{mismatch}$. It is a nonlinear filter that depends on patches or segments of the input (e.g., regions of an image) rather than individual pixels. Thus, we use convolution layers with kernel size as a $3 \times 3$  and padding size as $1$. Table ~\ref{tab: areawise} below shows five convolution layers, each followed by a Leaky Relu activation function ($\alpha = 0.02$) except for the last layer.

\begin{table}[ht]
\fontsize{8pt}{8pt}\selectfont
    \centering
    \begin{tabular}{c|c|c|c}
    \toprule
    \textbf{Layer Name} &\textbf{Output Channels} &\textbf{Kernel Size} & \hfil\textbf {Activation} \\\midrule
    Conv1 & 64 & $3 \times 3$ & LeakyReLU \\
    Conv2 & 64 & $3 \times 3$ & LeakyReLU \\
    Conv3 & 32 & $3 \times 3$ & LeakyReLU \\
    Conv4 & 32 & $3 \times 3$ & LeakyReLU \\
    Conv5 & 1 & $3 \times 3$ & - \\
    \bottomrule
    \end{tabular}
    \caption{\textbf{Neural areawise operator architecture.}}
    \label{tab: areawise}
\end{table}

\subsubsection{Fourier Neural Operator (FNO)} 
Our implementation of the FNO model is adapted from FNO2d in~\cite{li2020fourier}. Key hyperparameters are: $modes_1 = 12$; $modes_2 = 12$; $width = 24$.
We use a learning rate of $0.0001$ for training the FNO model.

\subsubsection{Parameterized physics model (Physics)}
The parameterized physics model is modified from our PBL model but substitutes the neural network modules with parameterized functions. Modifications include:
\begin{itemize}
    \item Substituting the pointwise neural network $g_{\text{thresh}}$ with a normalized projection ~\cite{zhou2014topology}:
    \begin{equation}
        \hat{H}_{\text{init}} = \frac{tanh(\alpha\eta)+tanh(\alpha(I_{\text{optical}}-\eta))}{tanh(\alpha\eta)+tanh(\alpha(1-\eta))},
    \end{equation}
    
    where $\alpha$ and $\eta$ are learnable single-value parameters. The function does not differ too much from sigmoid (normally used to represent the threshold behavior of photoresist) when $\alpha$ is large. We use it because, for a small $\alpha$, this function is better as it interpolates the whole interval [0,1], but the sigmoid function does not.
    
    \item Substituting the pointwise neural network $g_{\text{shrink}}$ with a linear mapping operation used in ~\cite{lang2022towards}: 
    
    \begin{equation}
        S = \frac{\kappa_{max} - \kappa_{min}}{max(I_\text{diffusion})-min(I_\text{diffusion})},        
    \end{equation}
    
    where $\kappa_{min}$ is the learnable single-value parameter and $\kappa_{max}$ is fixed to be 1.
    \item Substituting the areawise neural network $g_{\text{mismatch}}$ with a learnable single-value height bias $b$.
\end{itemize}
To summarize, in the parameterized physics model, they are $5$ learnable single-value parameters: $\alpha$, $\eta$, $\kappa_{min}$, $b$ and $\sigma_\text{diffusion}$, which is used to parameterize the kernel for the diffusion of reactants.
The learning rate for training the parameterized physics model is $0.01$.

\section{Additional Experimental Details}

\begin{figure}[h]
\centering
\includegraphics[width=.9\linewidth]{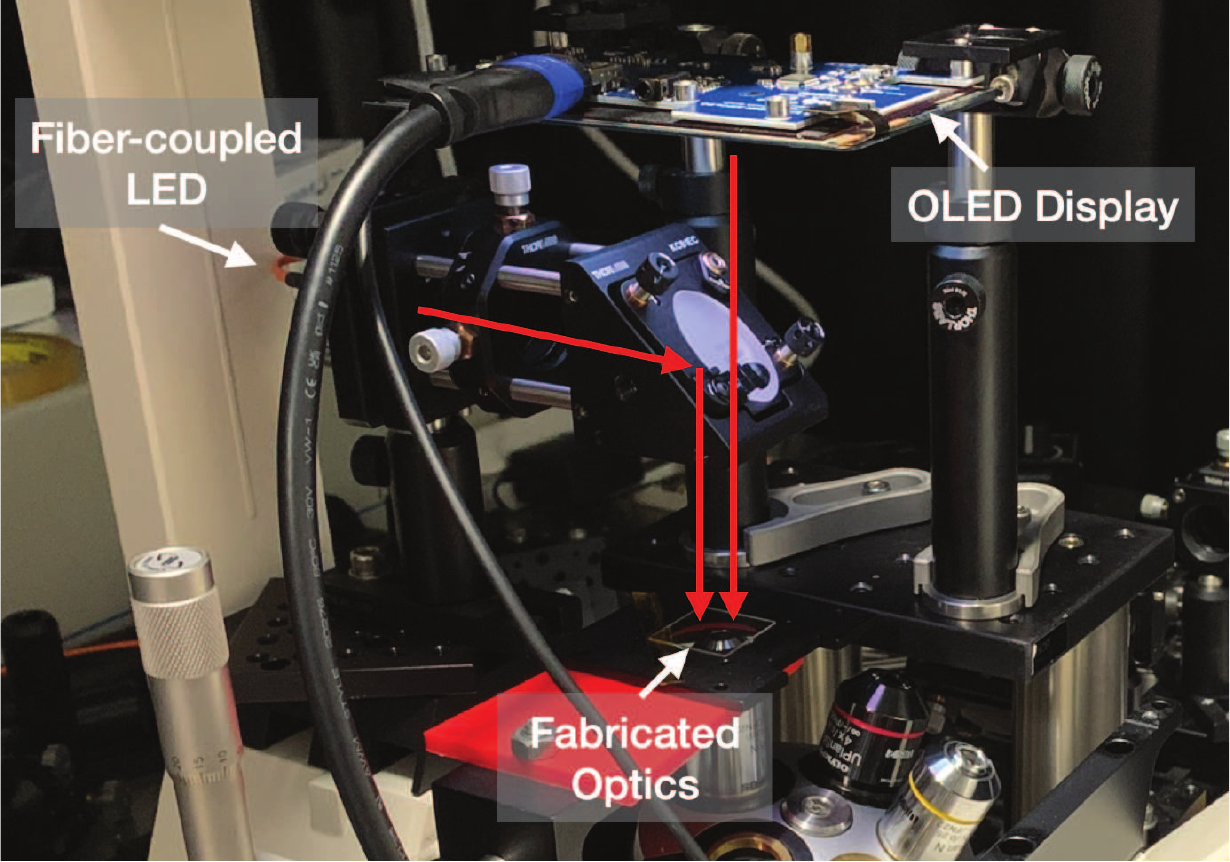}
\caption{\textbf{Experimental setup for assessing the performance of HOE and MDL.} The fiber-coupled LED calibrates the point spread function (PSF) in the MDL task or generates the holographic image in the HOE task. The OLED display here acts as the object for the MDL imaging task.
}
\label{fig: exp setup}
\end{figure}

The experimental setup in Fig.~\ref{fig: exp setup}, incorporates two illumination sources. For holographic imaging and the acquisition of the PSF, we utilize a fiber-coupled LED collimated to pass through a pinhole and then reflected by a mirror onto the fabricated optics. In MDL imaging, the fiber-coupled LED beam path is removed, and an OLED display serves as the object with dimensions of $36~mm$. The exposure times for capturing holographic and MDL images are 
$3~s$ and 
$5~s$, respectively. These extended durations are attributable to low illumination power and camera quantum efficiency. We also note the presence of dead pixels on the camera (CMOS, FLIR GS3-U3-32S4M-C), manifesting as random bright spots on the captured images, further compromising image quality.

\end{document}